# Thermo-optic refraction based switchable optical mode converter

Pritam P Shetty[1], Dmitrii N Maksimov[2,3], Mahalingam Babu[1], Sudhakara Reddy Bongu[4], Jayachandra Bingi[1, *]

[1] Bio-inspired Research and Development (BiRD) Laboratory, Photonic Devices and Sensors (PDS) Laboratory, Indian Institute of Information Technology Design and Manufacturing (IIITDM), Kancheepuram, Chennai 600127, India

[2] Kirensky Institute of Physics, Federal Research Center KSC SB RAS, 660036 Krasnoyarsk, Russia

[3] IRC SQC, Siberian Federal University, Krasnoyarsk 660041, Russia

[4] Department of Physics, Indian Institute of Technology Madras, Chennai 600036, India

* Corresponding author, email-id: bingi@iiitdm.ac.in

## Abstract

The temporally switchable optical mode conversion is crucial for optical communication and computing applications. This research demonstrates such optically switchable mode converter driven by thermo-optic refraction. The $MoS_2$ nanofluid is used as a medium where the thermal microlens is created by a focused laser beam (pump). The convective thermal plume generated above the focal point of the pump beam within the nanofluid acts as an astigmatic thermal lens. It is discovered that mode conversion of the Laguerre-Gaussian (LG) to the Hermite-Gaussian (HG) beam (vice versa) takes place upon passing through the thermal lens. The topological charge of the LG beam can be easily determined using the proposed mode converter. The mode transformation is explained theoretically as the Fourier components of the LG beam undergoing different optical paths while propagating through the convective plume.

## Introduction

The Laguerre-Gaussian (LG) beam is a type of an optical vortex beam with a doughnut shaped intensity distribution and helical phase structure. By imparting appropriate phase change to LG beams spatially, they readily transform into the corresponding Hermite-Gaussian (HG) beams [1]. A device for switching of spatial modes can be attractive for applications including space division multiplexing in free space optical communication [2], super resolution microscopy where structured light illumination is needed [3][4][5], and optical trapping and manipulation of nanoparticle[6][7]. Entangled photons carrying radial momentum have been explored for quantum technology[8]. Temporal mode switching is also used in quantum information science for processing of quantum data [9], [10]. For quantum communication applications, our proposed technique may be extended to generate spatio-temporal modes of single photon states for multiplexed data transmission.

A π/2 mode converter built using cylindrical lens can transform $LG_{p,\pm l}$ mode to $HG_{n,m}$ mode where radial index 'p' is min(n,m) and azimuthal index 'l' is (m-n), but such mode converter is not switchable [11]. Apertures are also used for topological charge detection, but lack mode switchability and capability to convert LG to HG mode and vice versa[12]. A spatial light modulator (SLM) can be used for switchable mode conversion by generating appropriate phase element but are very expensive and quality of beams depend on resolution of the SLM [13][14]. Fast switching of the mode conversion has been demonstrated in all fiber systems by the acousto-optic effect but have been demonstrated only for few higher order fiber core modes and may be affected by external vibration interference [15].

The optical Kerr effect is a phenomenon of an intense beam of light changing the refractive index of the pass-through medium proportional to the intensity. The Optical Kerr effect may result in self-focusing/defocusing, and spatial self-phase modulation (SSPM). The SSPM causes the beam transmitted through the material to self-interfere forming diffraction-like concentric ring patterns. The SSPM is observed in liquid crystals, dyes [16] and media containing 2D nanomaterials [17]. The SSPM effect has also been observed in 2D $MoS_2$ nanofluid and hence used in this research [18]–[20]. Further on, the optical Kerr effect caused by one beam used to affect phase of other beam passing through the same medium is called spatial cross phase modulation (SXPM) [21].

The application of the Kerr effect requires the use of optical materials with nonlinear susceptibility. In linear systems the effect of total internal reflection has been applied[22], [23] for beam transformation due to reflection of the Fourier components incident at the angle larger than critical. In this research switchable mode conversion from LG to HG mode and vice versa is demonstrated by SXPM in 2D $MoS_2$ nanofluid. The $MoS_2$ nanofluid is used as a medium where the thermal microlens is created by a focused laser beam (pump). The convective thermal plume generated above the focal point of the pump beam within the nanofluid acts as an astigmatic thermal lens. The effect of the lens leads to mode transformation due to the Fourier components of the LG beam undergoing different optical paths while propagating through the convective plume.

# Results and discussion

*Sample Preparation and characterization:*
2D $MoS_2$ nanoflakes are prepared by liquid phase exfoliation. The nanofluid consists of nanoflakes of $MoS_2$ dispersed in 3%w/w Polyvinylpyrrolidone (PVP) polymer solution. The material preparation is explained in our previous article [24]. UV-Vis absorption spectrum showed that $MoS_2$ nanofluid has broadband absorption in visible region (400-700nm). From scanning electron micrograph it is found that the nanoflakes had uniform size and average thickness of 153.5nm±23nm.

*Experimental results*

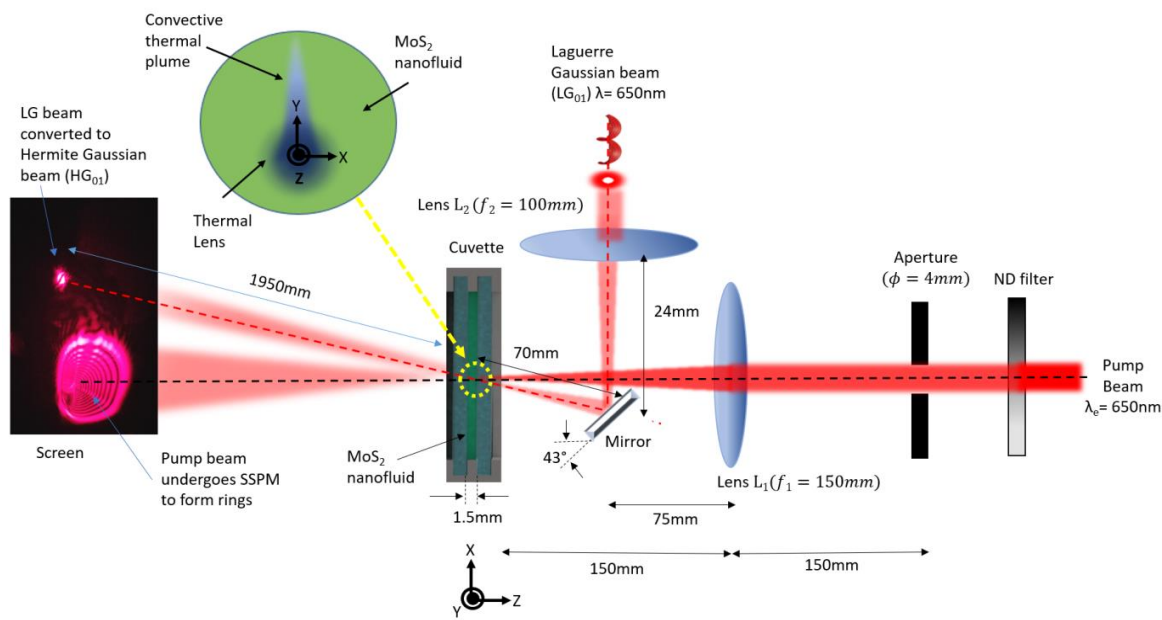

*Figure 1:The experimental setup for optically controlled mode converter*

Fig.1 shows the experimental setup for the optically controlled switchable mode converter. The pump beam is from a diode laser of 650nm wavelength and 60mW power and has the beam diameter of 4mm. The $LG_{0,1}$ beam entering $MoS_2$ nanofluid has beam size of ~146.7μm, the beam divergence of 14.6mrad and power of 147μW. The size of the thermal lens region can be tuned by controlling the power of pump beam using a variable neutral density filter. The pump beam is focused on $MoS_2$ nanofluid using lens L1. The suspended $MoS_2$ nanofluid absorbs the incident pump beam and creates localised heating. Due to the localised heating the refractive index of nanofluid changes in that region. The transmitted pump beam undergoes the SSPM and form symmetric concentric diffraction rings. Soon after, these concentric diffraction rings become asymmetric as shown in screen (Fig.1) due to the upward convective flow of the heated nanofluid. Details on thermal lens formation can be found in our previous article [24].

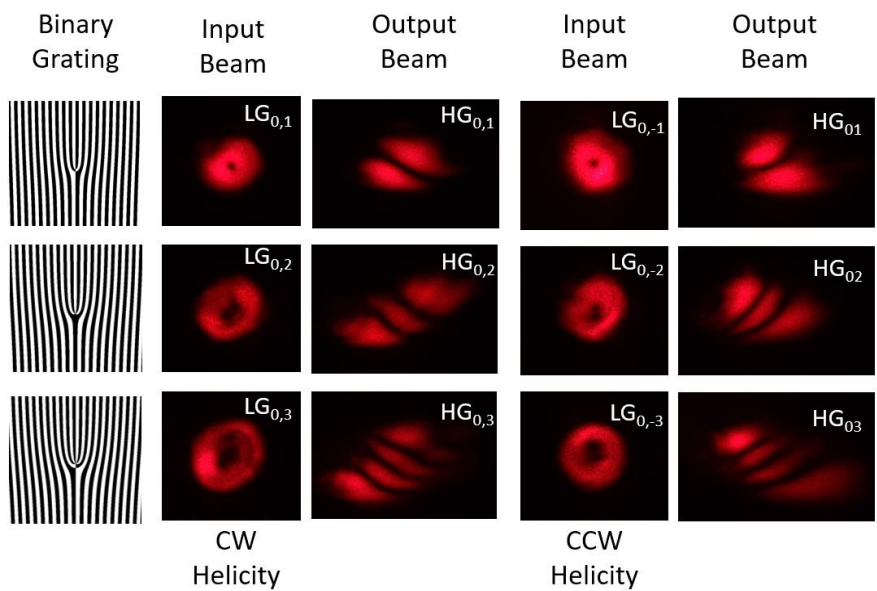

*Figure 2: The first column shows binary amplitude gratings used to generate LG beams. Second column shows input beams (Clockwise twisting) entering thermal lens region and third column shows output beams exiting thermal lens region. Similarly, Fourth and fifth column shows input beam (Counter-clockwise twisting) and output beam (HG beam).*

An inexpensive method of LG beam generation by forked binary amplitude gratings is used in this research. The size of gratings is 1cmx1cm and printed using laser printer on transparent overhead projector sheet. The first order diffraction from output of grating is an LG beam. Thus, beams of both clockwise and counter clockwise twist can be generated from a single fork binary grating. Fig.2 shows the input beam entering the thermal lens and the output beam exiting the thermal lens, respectively. The orientation and order of the output HG beams signified the twist direction and topological charge of the corresponding LG beam, respectively. The topological charge and twist of the LG beams up to order five were measured using this method. Higher topological charges can be calculated with equal efficiency using this technique with appropriate pump power adjustment. The decomposition of LG beam to corresponding HG beam using single cylindrical lens[25] and tilted convex lens[26], [27] has been reported in literature, but these methods does not allow control for dynamic optical mode switching. Moreover, to the best of our knowledge, free space optical mode conversion using thermally induced lens has never been reported. Vladimir et al have reported polychromatic OAM beam mode conversion using cylindrical lens, to achieve the same result using the proposed mode converter an optically transparent fluid like oils can be used[28].

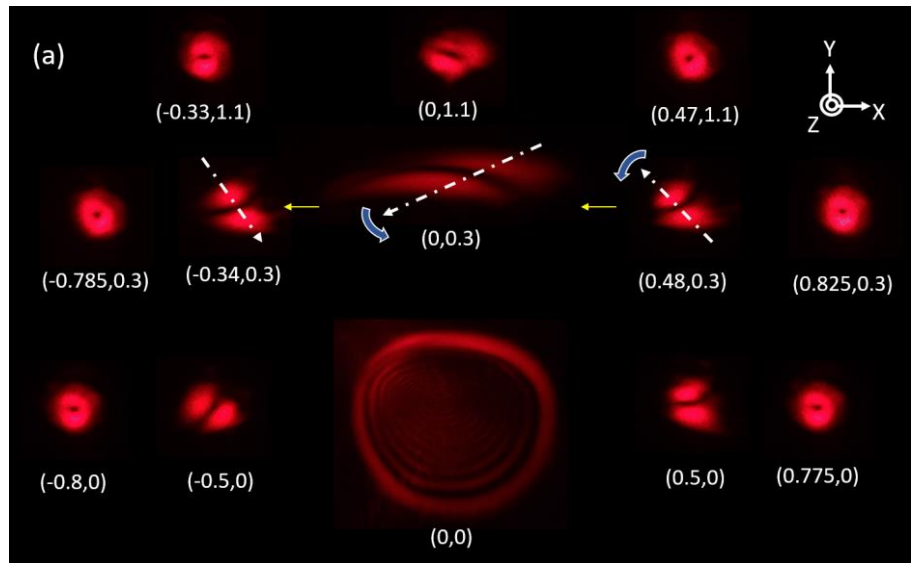

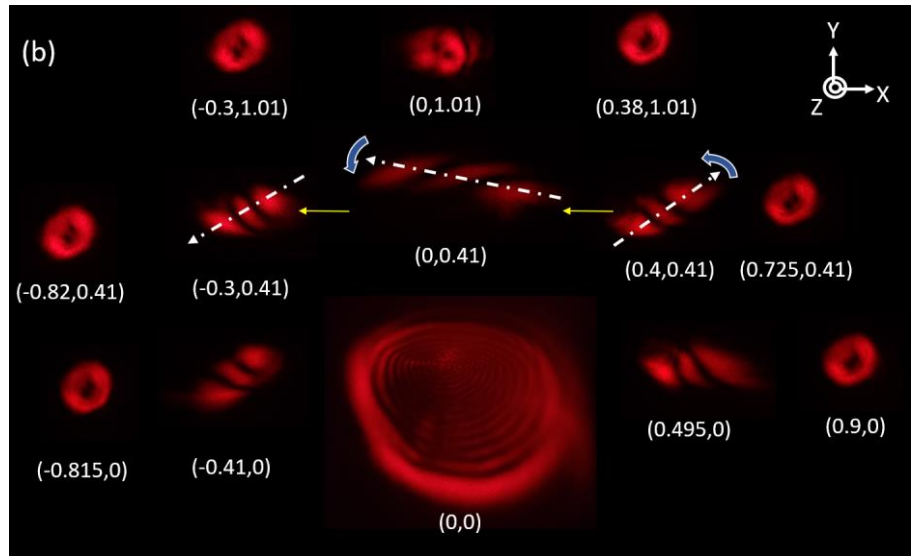

*Figure 3: (a) and(b) shows output beam patterns when $LG_{0,-1}$ and $LG_{0,2}$ modes are passed through thermal lens in nanofluid respectively. Co-ordinates show relative position (dimensions in mm) of input LG beam from origin(0,0). Here origin is beam axis of pump beam.*

The output beams were recorded by passing the input LG beams from right to left in the x-direction around thermal lens region (Fig.3). The centre of the thermal lens i.e. beam axis of pump beam is considered as the origin (0,0). When the LG beam passes through the origin, it turns into concentric ring like patterns similar to the pump beam undergoing the SSPM. When LG beam is moved away from origin along the x or y direction, the effect of mode conversion eventually disappears. This indicates the dependence of mode conversion on thermal lens shape. An interesting observation from Fig.3(a) is that if the location of the input $LG_{0,-1}$ beam is moved from (0.48, 0.3) to (-0.34, 0.3) position, the mode axis of the transformed output beam i.e., $HG_{0,1}$ rotates by 180° in anti-clockwise direction. Similarly, 180° rotation of the mode axis of output beam ($HG_{0,2}$) is observed for $LG_{0,2}$ beam. The reason for this phenomenon is a negative astigmatic thermal lens created by the pump beam. The heated convective thermal plume above origin has refractive index lower than that of the surrounding nanofluid and acts as a negative lens with the shape of bell as seen in Fig.1 (inset). Hence, the orientation of the output HG beam can give insights into the phase change experienced by LG beam.

Further, a theoretical analysis shows the same effect and is presented in the next section of the paper. The duration for complete mode switching takes ~0.6sec. This switching time can be further reduced by choosing nanofluid with a higher thermal conductivity and thermal coefficient of the refractive index.

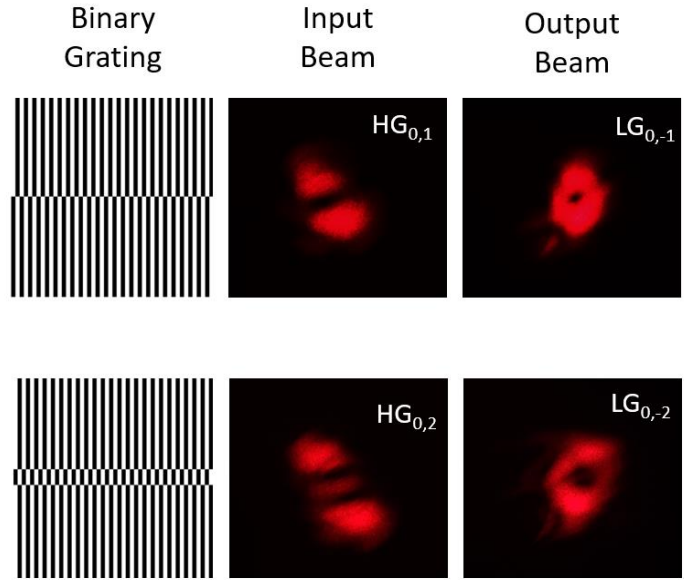

*Figure 4: First row shows binary amplitude grating used to generate HG beams. Second and third rows show input and output beams to/from thermal lens*

As shown in Fig.4 the HG beams can also be successfully converted into the corresponding LG beams by using the switchable mode converter. The HG beams were generated using binary amplitude grating as shown in Fig.4. The input HG beam is converted into the corresponding LG beam when it is inclined at ~25° to vertical. The Appropriate orientation HG beam is crucial for successful mode transformation.

*Theoretical results*

Similar to [29] the LG beam of radial index $p$ and azimuthal index $l$ with waist $w_0$ can be written in the following manner

$$E(\mathbf{r}) = (2f)^{2p+|l|}(-1)^{p+|l|}\left(\frac{2}{w^2}\right)^p \left(\frac{2(x+iy)}{w^2}\right)^l L_p^{|l|}\left(\frac{2(x^2+y^2)}{w^2}\right) E_0(\mathbf{r}) \qquad (1)$$

where $L_p^l(x)$ is the associated Laguerre polynomial, $z$ is the beam propagation axis, and $x$, $y$ are the coordinates in the beam Fourier (focal) plane,

$$f = \frac{1}{2} - \frac{iz}{kw_0^2}, \qquad (2)$$

$$w = w_0\sqrt{1 + \left(\frac{2z}{kw_0^2}\right)^2}, \qquad (3)$$

and

$$E_0(r) = \frac{1}{\pi}\frac{1}{(w_0^2+2iz/k)} \exp\left[ikz - i(k_x^2+k_y^2)z/2k - \frac{x^2+y^2}{(w_0^2+2iz/k)}\right], \qquad (4)$$

is the fundamental Gaussian mode. Alternatively, the LG beam can be written as its Fourier representation [30]

$$E(r) = \int_{-\infty}^{+\infty} \frac{dk_x dk_y}{(2\pi)^2} A_p^l(k_x,k_y) \exp[ik_x x + ik_y y + ikz - i(k_x^2+k_y^2)z/2k], \qquad (5)$$

where $A_p^l(k_x,k_y)$ is the Fourier amplitude,

$$A^l_p(k_x, k_y) = \left(\frac{2}{w_0^2}\right)^p (ik_x + k_y)^l \times L^l_p\left[\frac{(k_x^2+k_y^2)}{2w_0^2}\right] exp[-w_0^2(k_x^2+k_y^2)/4]. \quad (6)$$

The above equation gives us a key for understanding the beam transformation after passing through the thermo-optic lens. The effect is present at arbitrary angles of incidence. Thus, in contrast to [22], [23] the beam transformation cannot be attributed to the critical angle reflection of some Fourier components in Eq. (5). It is conjectured that the effect in this research can be explained through self-interference due to different phase shift for different Fourier components. The transmitted beam can be written as

$$E^{out}(r) = \int_{-\infty}^{+\infty} \frac{dk_x dk_y}{(2\pi)^2} exp[i\psi(k_x, k_y)] A^l_p(k_x, k_y) exp[ik_x x + ik_y y + ikz - i(k_x^2 + k_y^2)z/2k], \quad (7)$$

where the phase $\psi(k_x,k_y)$ is assumed to be dependent on the wave vector components in the Fourier plane. For the paraxial beams the range of $k_x, k_y$ is small in comparison to $k$, therefore $\psi(k_x,k_y)$ can be written as the Taylor expansion

$$\psi(k_x, k_y) = \psi_0 + \alpha_x \frac{k_x}{k} + \alpha_y \frac{k_y}{k} + \alpha_{xx}\left(\frac{k_x}{k}\right)^2 + \alpha_{yy}\left(\frac{k_y}{k}\right)^2 + \alpha_{xy}\frac{k_x k_y}{k^2}, \quad (8)$$

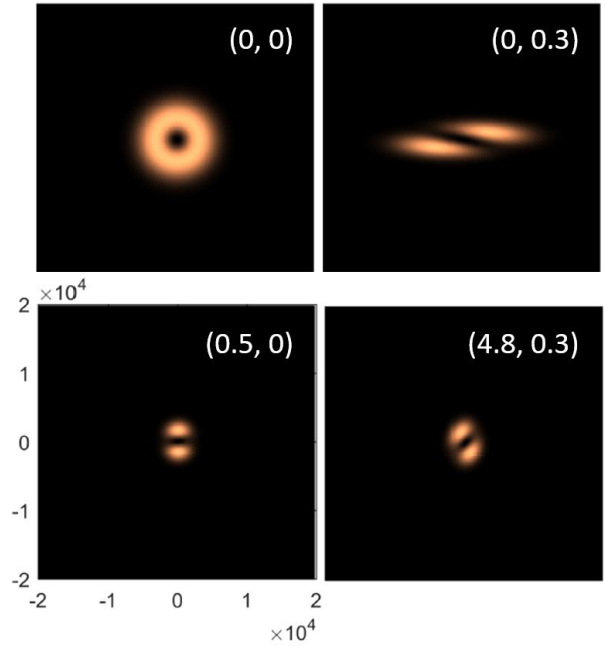

*Figure 5: Transformation of $LG_{0,-1}$ (counter clockwise beam).*

Notice that upon substitution of Eq. (8) in Eq. (7) $\alpha_x$, $\alpha_y$ can be removed by changing the coordinates in the Fourier plane $x^0 = x + \alpha_x$, $y^0 = y + \alpha_y$. Hence, the linear terms in Eq. (8) account for the beam shift, whereas $\psi_0$ is a global phase which can not affect the beam transformation. Thus, we conclude that it is $\alpha_{xx}$, $\alpha_{xy}$, $\alpha_{yy}$ that control the transformation effect. All three quantities are dependent on the orientation of the beam with respect to the thermal lens. However, they are independent of $l$, $n$, and $w_0$.

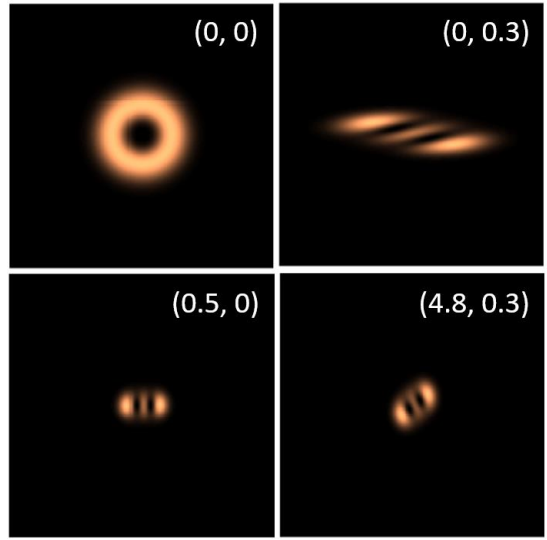

*Figure 6: Transformation of LG$_{0,2}$ (clockwise beam).*

In Table 1, the numerical values of the fitting parameters were obtained from the experimentally observed transformed beams patterns for the counter clockwise beam LG$_{(0,-1)}$. The corresponding beam profiles are plotted in Fig. 5. The obtained parameters for the clockwise beam LG$_{(0,2)}$ is used in next step. One can see in Fig. 6 that the three parameters result in the transformed beam profiles similar to those obtained in the experiment.

*Table 1: Fitting parameters obtained from LG$_{(0,-1)}$.*

| $\Delta x$ | 0 | 0 | 0.5 | 0.48 |
|---|---|---|---|---|
| $\Delta y$ | 0 | 0.3 | 0 | 0.3 |
| $\alpha_{xx}$ | 2×10$^6$ | 0.4×10$^6$ | 0.05×10$^6$ | 0.79×10$^6$ |
| $\alpha_{xy}$ | 2×10$^6$ | 3.8×10$^6$ | 0.02×10$^6$ | 0.04×10$^6$ |
| $\alpha_{yy}$ | 0 | 0 | 0.95 | 0.38×10$^6$ |

Hence, beam transformation can be explained as Fourier components of LG beam undergoing different optical path (phase shift) in the convective thermal plume. The beam transformation can be described by three fitting parameters that only depend on the vacuum wavelength and the orientation of the beam axis with respect to the convective plume. The parameters are independent of the beam extension, order, and azimuthal index (topological charge).

## Conclusion

In this research an optically switchable mode converter based on the thermo-optic refraction is demonstrated. The helicity and topological charge of the LG beam is easily measurable using this converter. Theoretically, it is found that mode transformation is a result of Fourier components on LG beam experiencing different optical paths while propagating through convective plume. These properties along with switchability of the demonstrated mode converter make it a potential tool for free space optical communication and quantum computing applications.

## Author contributions

**Pritam P Shetty**: Conceptualization, Methodology, Validation, Investigation, Visualization, Writing - Original Draft. **Dmitrii N Maksimov**: Theoretical validation, Formal analysis, Writing - Review &

Editing **Mahalingam Babu**: Investigation **Sudhakara Reddy Bongu:** Writing - Review & Editing **Jayachandra Bingi**: Conceptualization, Methodology, Writing - Review & Editing, Supervision.

## Declaration of Competing Interest

The authors declare that they have no known competing financial interests or personal relationships that could have appeared to influence the work reported in this paper.

## Acknowledgements

Authors acknowledge the funding support from DST India under INT/RUS/RFBR/P-262.

## Reference

[1] L. Allen, M. W. Beijersbergen, R. J. C. Spreeuw, and J. P. Woerdman, "Orbital angular momentum of light and the transformation of Laguerre-Gaussian laser modes," *Phys. Rev. A*, vol. 45, no. 11, pp. 8185–8189, 1992, doi: 10.1103/PhysRevA.45.8185.

[2] L. Li *et al.*, "High-Capacity Free-Space Optical Communications between a Ground Transmitter and a Ground Receiver via a UAV Using Multiplexing of Multiple Orbital-Angular-Momentum Beams," *Sci. Rep.*, vol. 7, no. 1, pp. 1–12, 2017, doi: 10.1038/s41598-017-17580-y.

[3] M. Ritsch-Marte, "Orbital angular momentum light in microscopy," *Philos. Trans. R. Soc. A Math. Phys. Eng. Sci.*, vol. 375, no. 2087, 2017, doi: 10.1098/rsta.2015.0437.

[4] J. Tang, J. Ren, and K. Y. Han, "Fluorescence imaging with tailored light," *Nanophotonics*, vol. 8, no. 12, pp. 2111–2128, 2019, doi: 10.1515/nanoph-2019-0227.

[5] T. A. Klar, E. Engel, and S. W. Hell, "Breaking Abbe's diffraction resolution limit in fluorescence microscopy with stimulated emission depletion beams of various shapes," *Phys. Rev. E - Stat. Physics, Plasmas, Fluids, Relat. Interdiscip. Top.*, vol. 64, no. 6, p. 9, 2001, doi: 10.1103/PhysRevE.64.066613.

[6] F. G. Mitri, "Cylindrical particle manipulation and negative spinning using a nonparaxial Hermite-Gaussian light-sheet beam," *Journal of Optics (United Kingdom)*, vol. 18, no. 10. 2016, doi: 10.1088/2040-8978/18/10/105402.

[7] M. Padgett and R. Bowman, "Tweezers with a twist," *Nat. Photonics*, vol. 5, no. 6, pp. 343–348, 2011, doi: 10.1038/nphoton.2011.81.

[8] L. Chen, T. Ma, X. Qiu, D. Zhang, W. Zhang, and R. W. Boyd, "Realization of the Einstein-Podolsky-Rosen Paradox Using Radial Position and Radial Momentum Variables," *Phys. Rev. Lett.*, vol. 123, no. 6, p. 60403, 2019, doi: 10.1103/PhysRevLett.123.060403.

[9] J. Ashby, V. Thiel, M. Allgaier, P. D'Ornellas, A. O. C. Davis, and B. J. Smith, "Temporal mode transformations by sequential time and frequency phase modulation for applications in quantum information science," Opt. Express, vol. 28, no. 25, p. 38376, 2020, doi: 10.1364/oe.410371.

[10] B. Brecht, D. V. Reddy, C. Silberhorn, and M. G. Raymer, "Photon temporal modes: A complete framework for quantum information science," *Phys. Rev. X*, vol. 5, no. 4, pp. 1–17, 2015, doi: 10.1103/PhysRevX.5.041017.

[11] M. W. Beijersbergen, L. Allen, H. E. L. O. van der Veen, and J. P. Woerdman, "Astigmatic laser mode converters and transfer of orbital angular momentum," *Opt.*

*Commun.*, vol. 96, no. 1–3, pp. 123–132, 1993, doi: 10.1016/0030-4018(93)90535-D.

[12] Y. Shen, X. Fu, and M. Gong, "Truncated triangular diffraction lattices and orbital-angular-momentum detection of vortex SU(2) geometric modes," *Opt. Express*, vol. 26, no. 20, p. 25545, 2018, doi: 10.1364/oe.26.025545.

[13] R. Chen, H. Zhou, M. Moretti, X. Wang, and J. Li, "Orbital Angular Momentum Waves: Generation, Detection, and Emerging Applications," *IEEE Commun. Surv. Tutorials*, vol. 22, no. 2, pp. 840–868, 2020, doi: 10.1109/COMST.2019.2952453.

[14] Y. Li, Y. Han, and Z. Cui, "Measuring the Topological Charge of Vortex Beams With Gradually Changing-Period Spiral Spoke Grating," *IEEE PHOTONICS Technol. Lett.*, vol. 32, no. 2, pp. 2019–2022, 2019.

[15] J. Lu *et al.*, "Dynamic mode-switchable optical vortex beams using acousto-optic mode converter," *Opt. Lett.*, vol. 43, no. 23, p. 5841, 2018, doi: 10.1364/ol.43.005841.

[16] H. Li *et al.*, "Off-resonant nonlinear optical refraction properties of azo dye doped nematic liquid crystals," *Opt. Mater. Express*, vol. 6, no. 2, p. 459, 2016, doi: 10.1364/ome.6.000459.

[17] L. Wu, X. Yuan, D. Ma, Y. Zhang, W. Huang, Y.Ge, Y. Song, Y. Xiang, J Li, H. Zhang,., "Recent Advances of Spatial Self-Phase Modulation in 2D Materials and Passive Photonic Device Applications," Small, vol. 16, no. 35. pp. 1–22, 2020, doi: 10.1002/smll.202002252.

[18] T. Neupane, B. Tabibi, and F. J. Seo, " Spatial self-phase modulation in WS 2 and MoS 2 atomic layers ," *Opt. Mater. Express*, vol. 10, no. 4, p. 831, 2020, doi: 10.1364/ome.380103.

[19] S. Xiao, B. Lv, L. Wu, M. Zhu, J. He, and S. Tao, "Dynamic self-diffraction in MoS_2 nanoflake solutions," *Opt. Express*, vol. 23, no. 5, p. 5875, 2015, doi: 10.1364/oe.23.005875.

[20] D. Sarkar, W. Liu, X. Xie, A. C. Anselmo, S. Mitragotri, and K. Banerjee, "MoS2 field-effect transistor for next-generation label-free biosensors," *ACS Nano*, vol. 8, no. 4, pp. 3992–4003, 2014, doi: 10.1021/nn5009148.

[21] Y. Jia, Y. Shan, L. Wu, X. Dai, D. Fan, and Y. Xiang, "Broadband nonlinear optical resonance and all-optical switching of liquid phase exfoliated tungsten diselenide," *Photonics Res.*, vol. 6, no. 11, p. 1040, 2018, doi: 10.1364/prj.6.001040.

[22] H. Okuda and H. Sasada, "Significant deformations and propagation variations of Laguerre-Gaussian beams reflected and transmitted at a dielectric interface," *J. Opt. Soc. Am. A*, vol. 25, no. 4, p. 881, 2008, doi: 10.1364/josaa.25.000881.

[23] H. Li, F. Honary, Z. Wu, and L. Bai, "Reflection and transmission of Laguerre-Gaussian beams in a dielectric slab," *J. Quant. Spectrosc. Radiat. Transf.*, vol. 195, pp. 35–43, 2017, doi: 10.1016/j.jqsrt.2016.12.001.

[24] P. P. Shetty, M. Babu, D. N. Maksimov, and J. Bingi, "Thermo-optic refraction in MoS2 medium for 'Normally on' all optical switch," *Opt. Mater. (Amst).*, vol. 112, p. 110777, Feb. 2021, doi: 10.1016/j.optmat.2020.110777.

[25] A. Y. Bekshaev, M. S. Soskin, and M. V. Vasnetsov, "Transformation of higher-order optical vortices upon focusing by an astigmatic lens," *Opt. Commun.*, vol. 241, no. 4–6, pp. 237–247, 2004, doi: 10.1016/j.optcom.2004.07.023.

[26] A. Y. Bekshaev and A. I. Karamoch, "Astigmatic telescopic transformation of a high-order optical vortex," *Opt. Commun.*, vol. 281, no. 23, pp. 5687–5696, 2008, doi: 10.1016/j.optcom.2008.09.017.

[27] V. V. Kotlyar, A. A. Kovalev, and A. P. Porfirev, "Astigmatic transforms of an optical vortex for measurement of its topological charge," *Appl. Opt.*, vol. 56, no. 14, p. 4095, 2017, doi: 10.1364/ao.56.004095.

[28] V. Denisenko *et al.*, "Determination of topological charges of polychromatic optical vortices," *Opt. Express*, vol. 17, no. 26, p. 23374, 2009, doi: 10.1364/oe.17.023374.

[29] J. Enderlein and F. Pampaloni, "Unified operator approach for deriving Hermite–Gaussian and Laguerre–Gaussian laser modes," *J. Opt. Soc. Am. A*, vol. 21, no. 8, p. 1553, 2004, doi: 10.1364/josaa.21.001553.

[30] K. N. Pichugin, D. N. Maksimov, and A. F. Sadreev, "Goos–Hänchen and Imbert–Fedorov shifts of higher-order Laguerre–Gaussian beams reflected from a dielectric slab," *J. Opt. Soc. Am. A*, vol. 35, no. 8, p. 1324, 2018, doi: 10.1364/josaa.35.001324.